# Zero-determinant strategies in iterated multi-strategy games

Jin-Li Guo (郭进利)

*Business School, University of Shanghai for Science and Technology, Shanghai 200093, China*

## Abstract

Self-serving, rational agents sometimes cooperate to their mutual benefit. The two-player iterated prisoner's dilemma game is a model for including the emergence of cooperation. It is generally believed that there is no simple ultimatum strategy which a player can control the return of the other participants. The recent discovery of the powerful class of zero-determinant strategies in the iterated prisoner's dilemma dramatically expands our understanding of the classic game by uncovering strategies that provide a unilateral advantage to sentient players pitted against unwitting opponents. However, strategies in the prisoner's dilemma game are only two strategies. Are there these results for general multi-strategy games? To address this question, the paper develops a theory for zero-determinant strategies for multi-strategy games, with any number of strategies. The analytical results exhibit a similar yet different scenario to the case of two-strategy games. Zero-determinant strategies in iterated Prisoner's Dilemma can be seen as degenerate case of our results. The results are also applied to the Snowdrift game, the Hawk–Dove game and the Chicken game.

## 1 Introduction

Although a game theory is initially emerged as a branch of mathematics, it covers almost every aspect of human interaction, especially including the mutual influence and interaction between human behavior, the interests of competition and cooperation between people, and the most successful applications in economics. One of the most used in game theory is a prisoner's dilemma, it is proposed by Tucker, and the study of the prisoner's dilemma involves mathematics, economics, political science, ethics, psychology, computer science and other fields. The prisoner's dilemma itself is well established as a way to study the emergence of cooperative behavior[1]. Each player is simultaneously offered two options: to cooperate or defect. If both players cooperate, they each receive the same payoff $R$; if both defect, they each receive a lower payoff $P$; if one player cooperates and the other defects, the defector receives the largest possible payoff $T$, and the cooperator the lowest possible payoff $S$. A dynamic iterated game is one of the newest directions in the studies of the game theory. Under the iterated game framework, the expected payoff of a player is determined by the others. It is difficult that an unilateral participant tries to find a simple optimal strategy. In PNAS, zero-determinant (ZD) strategies discovered by Press and Dyson [2] have attracted considerable attention [3-6]. They show that a player adopting zero-determinant strategies is able to pin the expected payoff of the opponents. In particular, a player $\alpha$ who is witting of these strategies can (i) deterministically set her opponent $\beta$'s score, independently of his strategy or response, or (ii) enforce an extortionate linear relation between her and his scores [2]. Despite being not consistent with our intuition, it describes a beautiful outlook for the dynamic iterated game and also causes the much attention of many scientists [3-7]. Szolnoki and Perc [3] studied the evolution of cooperation in the spatial prisoner's dilemma game, where besides unconditional cooperation and defection, tit-for-tat, win-stay-lose-shift and extortion are the five competing strategies. To explore the performance of ZD strategies against humans, Hilbe et al.[4] have designed an economic experiment in which participants were matched either with an extortioner or with a generous co-player. They show, although extortioners succeeded against each of their human opponents, extortion resulted in lower payoffs than generosity. Stewart and Plotkin [5] explored the evolutionary prospects for ZD strategies in the iterated prisoner's dilemma. Hilbe et al. [8] studied zero-determinant alliances in multiplayer social dilemmas. Zero-determinant strategies were also generalized for the iterated public goods game [9] and all symmetric 2x2 games [10].

As mentioned above, however, these games are two-strategy games. One aim of this paper is to design a framework for zero-determinant strategies in iterated multi-strategy games. It is found by surprise that the ZD strategies still exist for a player with many strategies in two-player iterated games. In the first place we develop ZD strategies in iterated multi-strategy games and give a feasible condition of ZD strategies. In the second place we study the mechanisms of zero-determinant strategies in symmetric games and the mischief or the extortion of a player. Third, the results are applied to the Snowdrift game, the Hawk–Dove game and the Chicken game.

## 2 Framework of zero-determinant strategies

A multi-strategy game has the following two characteristics：

(1) Let $\alpha$ and $\beta$ denote player 1 and player 2, respectively. Both players have only a finite number of strategies. Assuming that there are $n$ strategies for player $\alpha$, $m$ ($1 \leq m \leq n$) strategies for player $\beta$. Strategy sets are respectively represented as follows: $S_\alpha = \{\alpha_1, \alpha_2, \alpha_3, \cdots, \alpha_n\}$, $S_\beta = \{\beta_1, \beta_2, \beta_3, \cdots, \beta_m\}$.

(2) In each round game, if player $\alpha$ and player $\beta$ use strategy $\alpha_i$ and strategy $\beta_j$, respectively, then a game $(\alpha_i, \beta_j)$ is formed. Suppose that player $\alpha$ receives payoff $a_{ij}$ and player $\beta$ receives payoff $b_{ji}$, then the payoff matrix of player $\alpha$ is $A = [a_{ij}]_{n \times m}$, the payoff matrix of player $\beta$ is $B = [b_{ij}]_{m \times n}$.

The data of a finite two-person game can be summarized by two matrices. Two-person games with finitely many choices, like the one above, are also called matrix games since they can be represented by two matrixes. Usually, these matrices are written as one matrix with two numbers at each position. Therefore, such games are often called 'bimatrix games'. The formal definition is as follows. A bimatrix game is a pair of $n \times m$ matrices $(A, B)$, where

$$(A, B) = (a_{ij}, b_{ij})_{n \times m} = \begin{pmatrix} (a_{11}, b_{11}) & (a_{12}, b_{12}) & \cdots & (a_{1m}, b_{1m}) \\ (a_{21}, b_{21}) & (a_{22}, b_{22}) & \cdots & (a_{2m}, b_{2m}) \\ \vdots & \vdots & \cdots & \vdots \\ (a_{n1}, b_{n1}) & (a_{n2}, b_{n2}) & \cdots & (a_{nm}, b_{nm}) \end{pmatrix}. \quad (1)$$

In iterated games, for player $\alpha$ the possible outcome of each stage game can be represented as: $(\alpha_i, \beta_j)_{n \times m}$.

For player $\alpha$, the conditional probability that in next game strategy $\alpha_k$ occurs given that current game $(\alpha_i, \beta_j)$ has occurred is denoted by $p^{(k)}_{\alpha_i \beta_j}$. Therefore, the conditional probability vector is formed as follows

$$p^{(k)} = (p^{(k)}_{\alpha_1 \beta_1}, p^{(k)}_{\alpha_1 \beta_2}, \cdots, p^{(k)}_{\alpha_1 \beta_m}, p^{(k)}_{\alpha_2 \beta_1}, p^{(k)}_{\alpha_2 \beta_2}, \cdots, p^{(k)}_{\alpha_2 \beta_m}, \cdots, p^{(k)}_{\alpha_n \beta_1}, p^{(k)}_{\alpha_n \beta_2}, \cdots, p^{(k)}_{\alpha_n \beta_m}),$$
$$k = 1, 2, \cdots, n \quad (2)$$

For player $\beta$ the possible outcome of each stage game can be represented as: $(\beta_j, \alpha_i)_{m \times n}$.

For player $\beta$, the conditional probability that in next game strategy $\beta_k$ occurs given that current game $(\beta_j, \alpha_i)$ has occurred is denoted by $q^{(k)}_{\beta_j \alpha_i}$. Therefore, the conditional probability matrix is formed as follows

$$q^{(k)} = (q^{(k)}_{\beta_1 \alpha_1}, q^{(k)}_{\beta_1 \alpha_2}, \cdots, q^{(k)}_{\beta_1 \alpha_n}, q^{(k)}_{\beta_2 \alpha_1}, q^{(k)}_{\beta_2 \alpha_2}, \cdots, q^{(k)}_{\beta_2 \alpha_n}, \cdots, q^{(k)}_{\beta_m \alpha_1}, q^{(k)}_{\beta_m \alpha_2}, \cdots, q^{(k)}_{\beta_m \alpha_n}),$$
$$k = 1, 2, \cdots, m \quad (3)$$

We let $p$ and $q$ denote $p^{(1)}$ and $q^{(1)}$, respectively. $p^{(k)}$ and $q^{(k)}$ imply a Markov matrix whose stationary probability vector $v$, combined with the respective payoff matrices, yields an expected outcome for each player. With rows and columns of the matrix in $\alpha$'s order, the Markov transition matrix $P(p, q)$ from one move to the next is shown in Fig.1.

$$P = \begin{pmatrix} p^{(1)}_{\alpha_1\beta_1}q^{(1)}_{\beta_1\alpha_1} & p^{(1)}_{\alpha_1\beta_1}q^{(2)}_{\beta_1\alpha_1} & \cdots & p^{(1)}_{\alpha_1\beta_1}q^{(m)}_{\beta_1\alpha_1} & p^{(2)}_{\alpha_1\beta_1}q^{(1)}_{\beta_1\alpha_1} & p^{(2)}_{\alpha_1\beta_1}q^{(2)}_{\beta_1\alpha_1} & \cdots & p^{(2)}_{\alpha_1\beta1}q^{(m)}_{\beta_1\alpha_1} & \cdots p^{(n)}_{\alpha_1\beta_1}q^{(1)}_{\beta_1\alpha_1} & \cdots p^{(n)}_{\alpha_1\beta_1}q^{(m)}_{\beta_1\alpha_1} \\ p^{(1)}_{\alpha_2\beta_1}q^{(1)}_{\beta_1\alpha_2} & p^{(1)}_{\alpha_2\beta_1}q^{(2)}_{\beta_1\alpha_2} & \cdots & p^{(1)}_{\alpha_2\beta_1}q^{(m)}_{\beta_1\alpha_2} & p^{(2)}_{\alpha_2\beta_1}q^{(1)}_{\beta_1\alpha_2} & p^{(2)}_{\alpha_2\beta_1}q^{(2)}_{\beta_1\alpha_2} & \cdots & p^{(2)}_{\alpha_2\beta_1}q^{(m)}_{\beta_1\alpha_2} & \cdots p^{(n)}_{\alpha_2\beta_1}q^{(1)}_{\beta_1\alpha_2} & \cdots p^{(n)}_{\alpha_2\beta_1}q^{(m)}_{\beta_1\alpha_2} \\ \vdots & \vdots & \cdots & \vdots & \vdots & \vdots & \cdots & \vdots & \vdots & \vdots \\ p^{(1)}_{\alpha_n\beta_1}q^{(1)}_{\beta_1\alpha_n} & p^{(1)}_{\alpha_n\beta_1}q^{(2)}_{\beta_1\alpha_n} & \cdots & p^{(1)}_{\alpha_n\beta_1}q^{(m)}_{\beta_1\alpha_n} & p^{(2)}_{\alpha_n\beta_1}q^{(1)}_{\beta_1\alpha_n} & p^{(2)}_{\alpha_n\beta_1}q^{(2)}_{\beta_1\alpha_n} & \cdots & p^{(2)}_{\alpha_n\beta_1}q^{(m)}_{\beta_1\alpha_n} & \cdots p^{(n)}_{\alpha_n\beta_1}q^{(1)}_{\beta_1\alpha_n} & \cdots p^{(n)}_{\alpha_n\beta_1}q^{(m)}_{\beta_1\alpha_n} \\ p^{(1)}_{\alpha_1\beta_2}q^{(1)}_{\beta_2\alpha_1} & p^{(1)}_{\alpha_1\beta_2}q^{(2)}_{\beta_2\alpha_1} & \cdots & p^{(1)}_{\alpha_1\beta_2}q^{(m)}_{\beta_2\alpha_1} & p^{(2)}_{\alpha_1\beta_2}q^{(1)}_{\beta_2\alpha_1} & p^{(2)}_{\alpha_1\beta_2}q^{(2)}_{\beta_2\alpha_1} & \cdots & p^{(2)}_{\alpha_1\beta_2}q^{(m)}_{\beta_2\alpha_1} & \cdots p^{(n)}_{\alpha_1\beta_2}q^{(1)}_{\beta_2\alpha_1} & \cdots p^{(n)}_{\alpha_1\beta_2}q^{(m)}_{\beta_2\alpha_1} \\ p^{(1)}_{\alpha_2\beta_2}q^{(1)}_{\beta_2\alpha_2} & p^{(1)}_{\alpha_2\beta_2}q^{(2)}_{\beta_2\alpha_2} & \cdots & p^{(1)}_{\alpha_2\beta_2}q^{(m)}_{\beta_2\alpha_2} & p^{(2)}_{\alpha_2\beta_2}q^{(1)}_{\beta_2\alpha_2} & p^{(2)}_{\alpha_2\beta_2}q^{(2)}_{\beta_2\alpha_2} & \cdots & p^{(2)}_{\alpha_2\beta_2}q^{(m)}_{\beta_2\alpha_2} & \cdots p^{(n)}_{\alpha_2\beta_2}q^{(1)}_{\beta_2\alpha_2} & \cdots p^{(n)}_{\alpha_2\beta_2}q^{(m)}_{\beta_2\alpha_2} \\ \vdots & \vdots & \cdots & \vdots & \vdots & \vdots & \cdots & \vdots & \vdots & \vdots \\ p^{(1)}_{\alpha_n\beta_2}q^{(1)}_{\beta_2\alpha_n} & p^{(1)}_{\alpha_n\beta_2}q^{(2)}_{\beta_2\alpha_n} & \cdots & p^{(1)}_{\alpha_n\beta_2}q^{(m)}_{\beta_2\alpha_n} & p^{(2)}_{\alpha_n\beta_2}q^{(1)}_{\beta_2\alpha_n} & p^{(2)}_{\alpha_n\beta_2}q^{(2)}_{\beta_2\alpha_n} & \cdots & p^{(2)}_{\alpha_n\beta_2}q^{(m)}_{\beta_2\alpha_n} & \cdots p^{(n)}_{\alpha_n\beta_2}q^{(1)}_{\beta_2\alpha_n} & \cdots p^{(n)}_{\alpha_n\beta_2}q^{(m)}_{\beta_2\alpha_n} \\ \vdots & \vdots & \cdots & \vdots & \vdots & \vdots & \cdots & \vdots & \cdots & \vdots \\ p^{(1)}_{\alpha_1\beta_m}q^{(1)}_{\beta_m\alpha_1} & p^{(1)}_{\alpha_1\beta_m}q^{(2)}_{\beta_m\alpha_1} & \cdots & p^{(1)}_{\alpha_1\beta_m}q^{(m)}_{\beta_m\alpha_1} & p^{(2)}_{\alpha_1\beta_m}q^{(1)}_{\beta_m\alpha_1} & p^{(2)}_{\alpha_1\beta_m}q^{(2)}_{\beta_m\alpha_1} & \cdots & p^{(2)}_{\alpha_1\beta_m}q^{(m)}_{\beta_m\alpha_1} & \cdots p^{(n)}_{\alpha_1\beta_m}q^{(1)}_{\beta_m\alpha_1} & \cdots p^{(n)}_{\alpha_1\beta_m}q^{(m)}_{\beta_m\alpha_1} \\ \vdots & \vdots & \cdots & \vdots & \vdots & \vdots & \cdots & \vdots & \vdots & \vdots \\ p^{(1)}_{\alpha_n\beta_m}q^{(1)}_{\beta_m\alpha_n} & p^{(1)}_{\alpha_n\beta_m}q^{(2)}_{\beta_m\alpha_n} & \cdots & p^{(1)}_{\alpha_n\beta_m}q^{(m)}_{\beta_m\alpha_n} & p^{(2)}_{\alpha_n\beta_m}q^{(1)}_{\beta_m\alpha_n} & p^{(2)}_{\alpha_n\beta_m}q^{(2)}_{\beta_m\alpha_n} & \cdots & p^{(2)}_{\alpha_n\beta_m}q^{(m)}_{\beta_m\alpha_n} & \cdots p^{(n)}_{\alpha_n\beta_m}q^{(1)}_{\beta_m\alpha_n} & \cdots p^{(n)}_{\alpha_n\beta_m}q^{(m)}_{\beta_m\alpha_n} \end{pmatrix}$$

**Fig. 1.**  The Markov transition matrix $P(p,q)$.

Because $P$ has a unit eigenvalue, the matrix $P' \equiv P - I$ is singular, with thus zero determinant. The stationary vector $v$ of the Markov matrix, or any vector proportional to it, satisfies

$$vP = v \qquad \text{or} \qquad vP' = 0. \tag{4}$$

The adjugate matrix of $P'$ is as follows

$$Adj(P') = \begin{pmatrix} P_{11} & P_{21} & \cdots & P_{nm1} \\ P_{12} & P_{22} & \cdots & P_{nm2} \\ \vdots & \vdots & \cdots & \vdots \\ P_{1nm} & P_{2nm} & \cdots & P_{nmnm} \end{pmatrix}, \tag{5}$$

$|P'| \equiv |P - I| = 0$ implying that

$$Adj(P')P' = |P'|I = 0. \tag{6}$$

Based on the properties of the adjugate matrix and the stationary probability, every row of $Adj(P')$ is proportional to *v*. Choosing the last row, we see that the components of *v* are (up to a sign) the determinants of the *nm* × *nm* matrices formed from the first *nm*-1 columns of $P'$, leaving out each one of the *nm* rows in turn. These determinants are unchanged if the first column of $P'$ is added into the second column, and third column is added into the first column.

The result of these manipulations is a formula for the dot product of an arbitrary *nm*-vector *f* with the stationary vector *v* of the Markov matrix, $v \cdot f \equiv D(p,q,f)$, where $D(p,q,f)$ is the following determinant (see Fig.2).

$$\begin{vmatrix} q_{\beta_1\alpha_1}^{(1)}-1 & p_{\alpha_1\beta_1}^{(1)}-1 & p_{\alpha_1\beta_1}^{(1)}q_{\beta_1\alpha_1}^{(3)} & \cdots & p_{\alpha_1\beta_1}^{(2)}q_{\beta_1\alpha_1}^{(1)} & p_{\alpha_1\beta_1}^{(2)}q_{\beta_1\alpha_1}^{(2)} & \cdots & p_{\alpha_1\beta_1}^{(2)}q_{\beta_1\alpha_1}^{(m)} & \cdots & p_{\alpha_1\beta_1}^{(n)}q_{\beta_1\alpha_1}^{(m-1)} & f_1 \\ q_{\beta_1\alpha_2}^{(1)} & p_{\alpha_2\beta_1}^{(1)}-1 & p_{\alpha_2\beta_1}^{(1)}q_{\beta_1\alpha_2}^{(3)} & \cdots & p_{\alpha_2\beta_1}^{(2)}q_{\beta_1\alpha_2}^{(1)} & p_{\alpha_2\beta_1}^{(2)}q_{\beta_1\alpha_2}^{(2)} & \cdots & p_{\alpha_2\beta_1}^{(2)}q_{\beta_1\alpha_2}^{(m)} & \cdots & p_{\alpha_2\beta_1}^{(n)}q_{\beta_1\alpha_2}^{(m-1)} & f_2 \\ \vdots & \vdots & \vdots & \vdots & \vdots & \vdots & \cdots & \vdots & \vdots & \vdots & \vdots \\ q_{\beta_1\alpha_{m-1}}^{(1)} & p_{\alpha_{m-1}\beta_1}^{(1)}-1 & p_{\alpha_{m-1}\beta_1}^{(1)}q_{\beta_1\alpha_{m-1}}^{(3)} & \cdots & p_{\alpha_{m-1}\beta_1}^{(2)}q_{\beta_1\alpha_{m-1}}^{(1)} & p_{\alpha_{m-1}\beta_1}^{(2)}q_{\beta_1\alpha_{m-1}}^{(2)} & \cdots & p_{\alpha_{m-1}\beta_1}^{(2)}q_{\beta_1\alpha_{m-1}}^{(m)} & \cdots & p_{\alpha_{m-1}\beta_1}^{(n)}q_{\beta_1\alpha_{m-1}}^{(m-1)} & f_{m-1} \\ q_{\beta_1\alpha_m}^{(1)} & p_{\alpha_m\beta_1}^{(1)}-1 & p_{\alpha_m\beta_1}^{(1)}q_{\beta_1\alpha_m}^{(3)} & \cdots & p_{\alpha_m\beta_1}^{(2)}q_{\beta_1\alpha_m}^{(1)} & p_{\alpha_m\beta_1}^{(2)}q_{\beta_1\alpha_m}^{(2)} & \cdots & p_{\alpha_m\beta_1}^{(2)}q_{\beta_1\alpha_m}^{(m)} & \cdots & p_{\alpha_m\beta_1}^{(n)}q_{\beta_1\alpha_m}^{(m-1)} & f_m \\ q_{\beta_1\alpha_{m+1}}^{(1)}-1 & p_{\alpha_{m+1}\beta_1}^{(1)} & p_{\alpha_{m+1}\beta_1}^{(1)}q_{\beta_1\alpha_{m+1}}^{(3)} & \cdots & p_{\alpha_{m+1}\beta_1}^{(2)}q_{\beta_1\alpha_{m+1}}^{(1)} & p_{\alpha_{m+1}\beta_1}^{(2)}q_{\beta_1\alpha_{m+1}}^{(2)} & \cdots & p_{\alpha_{m+1}\beta_1}^{(2)}q_{\beta_1\alpha_{m+1}}^{(m)} & \cdots & p_{\alpha_{m+1}\beta_1}^{(n)}q_{\beta_1\alpha_{m+1}}^{(m-1)} & f_{m+1} \\ q_{\beta_1\alpha_{m+2}}^{(1)} & p_{\alpha_{m+2}\beta_1}^{(1)} & p_{\alpha_{m+2}\beta_1}^{(1)}q_{\beta_1\alpha_{m+2}}^{(3)}-1 & \vdots & p_{\alpha_{m+2}\beta_1}^{(2)}q_{\beta_1\alpha_{m+2}}^{(1)} & p_{\alpha_{m+2}\beta_1}^{(2)}q_{\beta_1\alpha_{m+2}}^{(2)} & \cdots & p_{\alpha_{m+2}\beta_1}^{(2)}q_{\beta_1\alpha_{m+2}}^{(m)} & \vdots & p_{\alpha_{m+2}\beta_1}^{(n)}q_{\beta_1\alpha_{m+2}}^{(m-1)} & f_{m+2} \\ \vdots & \vdots & \vdots & \cdots & \vdots & \vdots & \cdots & \vdots & \cdots & \vdots & \vdots \\ q_{\beta_2\alpha_1}^{(1)} & p_{\alpha_1\beta_2}^{(1)} & p_{\alpha_1\beta_2}^{(1)}q_{\beta_2\alpha_1}^{(3)} & \cdots & p_{\alpha_1\beta_2}^{(2)}q_{\beta_2\alpha_1}^{(1)} & p_{\alpha_1\beta_2}^{(2)}q_{\beta_2\alpha_1}^{(2)} & \cdots & p_{\alpha_1\beta_2}^{(2)}q_{\beta_2\alpha_1}^{(m)} & \cdots & p_{\alpha_1\beta_2}^{(n)}q_{\beta_2\alpha_1}^{(m-1)} & f_{n+1} \\ q_{\beta_2\alpha_2}^{(1)} & p_{\alpha_2\beta_2}^{(1)} & p_{\alpha_2\beta_2}^{(1)}q_{\beta_2\alpha_2}^{(3)} & \cdots & p_{\alpha_2\beta_2}^{(2)}q_{\beta_2\alpha_2}^{(1)} & p_{\alpha_2\beta_2}^{(2)}q_{\beta_2\alpha_2}^{(2)} & \cdots & p_{\alpha_2\beta_2}^{(2)}q_{\beta_2\alpha_2}^{(m)} & \cdots & p_{\alpha_2\beta_2}^{(n)}q_{\beta_2\alpha_2}^{(m-1)} & f_{n+2} \\ \vdots & \vdots & \cdots & \vdots & \vdots & \vdots & \cdots & \vdots & \cdots & \vdots & \vdots \\ q_{\beta_2\alpha_{m+1}}^{(1)}-1 & p_{\alpha_{m+1}\beta_2}^{(1)} & p_{\alpha_{m+1}\beta_2}^{(1)}q_{\beta_2\alpha_{m+1}}^{(3)} & \cdots & p_{\alpha_{m+1}\beta_2}^{(2)}q_{\beta_2\alpha_{m+1}}^{(1)} & p_{\alpha_{m+1}\beta_2}^{(2)}q_{\beta_2\alpha_{m+1}}^{(2)} & \cdots & p_{\alpha_{m+1}\beta_2}^{(2)}q_{\beta_2\alpha_{m+1}}^{(m)} & \cdots & p_{\alpha_{m+1}\beta_2}^{(n)}q_{\beta_2\alpha_{m+1}}^{(m-1)} & f_{n+m} \\ \vdots & \vdots & \cdots & \vdots & \vdots & \vdots & \cdots & \vdots & \cdots & \vdots & \vdots \\ q_{\beta_m\alpha_1}^{(1)} & p_{\alpha_1\beta_m}^{(1)} & p_{\alpha_1\beta_m}^{(1)}q_{\beta_m\alpha_1}^{(3)} & \cdots & p_{\alpha_1\beta_m}^{(2)}q_{\beta_m\alpha_1}^{(1)} & p_{\alpha_1\beta_m}^{(2)}q_{\beta_m\alpha_1}^{(2)} & \cdots & p_{\alpha_1\beta_m}^{(2)}q_{\beta_m\alpha_1}^{(m)} & \cdots & p_{\alpha_1\beta_m}^{(n)}q_{\beta_m\alpha_1}^{(m-1)} & f_{nm-1} \\ \vdots & \vdots & \cdots & \vdots & \vdots & \vdots & \cdots & \vdots & \cdots & \vdots & \vdots \\ q_{\beta_m\alpha_n}^{(1)} & p_{\alpha_n\beta_m}^{(1)} & p_{\alpha_n\beta_m}^{(1)}q_{\beta_m\alpha_n}^{(3)} & \cdots & p_{\alpha_n\beta_m}^{(2)}q_{\beta_m\alpha_n}^{(1)} & p_{\alpha_n\beta_m}^{(2)}q_{\beta_m\alpha_n}^{(2)} & \cdots & p_{\alpha_n\beta_m}^{(2)}q_{\beta_m\alpha_n}^{(m)} & \cdots & p_{\alpha_n\beta_m}^{(n)}q_{\beta_m\alpha_n}^{(m-1)} & f_{nm} \end{vmatrix}$$

**Fig.2.** Determinant $D(p,q,f)$

This result follows from expanding the determinant by minors on its *nm*-th column and noting that the *nm*-1 × *nm*-1 determinants multiplying each $f_i$ are just the ones described above. What is noteworthy about this formula for $v \cdot f$ is that it is a determinant whose second column,

$$\bar{p} = (p_{\alpha_1\beta_1}^{(1)}-1, \cdots, p_{\alpha_m\beta_1}^{(1)}-1, p_{\alpha_{m+1}\beta_1}^{(1)}, \cdots, p_{\alpha_n\beta_1}^{(1)}, p_{\alpha_1\beta_2}^{(1)}, \cdots, p_{\alpha_n\beta_2}^{(1)}, \cdots, p_{\alpha_1\beta_m}^{(1)}, \cdots p_{\alpha_1\beta_m}^{(1)}), \quad (7)$$

is solely under the control of $\alpha$; whose first column,

$$\bar{q} = (q_{\beta_1\alpha_1}^{(1)}-1, q_{\beta_1\alpha_2}^{(1)}, \cdots, q_{\beta_1\alpha_m}^{(1)} q_{\beta_1\alpha_{m+1}}^{(1)}-1, q_{\beta_1\alpha_{m+2}}^{(1)}, \cdots, q_{\beta_1\alpha_n}^{(1)}, q_{\beta_2\alpha_1}^{(1)}, \cdots, q_{\beta_2\alpha_{m+1}}^{(1)}-1, \cdots, q_{\beta_2\alpha_n}^{(1)}, \cdots q_{\beta_m\alpha_n}^{(1)}). \quad (8)$$

is solely under the control of $\beta$; and whose *nm*-th column is simply *f*.

We rewrite the payoff matrix player $\alpha$ as follows vector form

$$\omega_\alpha = (a_{11}, a_{12}, \cdots a_{1m}, a_{21}, a_{22}, \cdots, a_{2m}, \cdots, a_{n1}, a_{n2}, \cdots, a_{nm}), \quad (9)$$

We also rewrite the payoff matrix player $\beta$ as follows vector form

$$\omega_\beta = (b_{11}, b_{12}, \cdots b_{1n}, b_{21}, b_{22}, \cdots, b_{2n}, \cdots, b_{m1}, b_{m2} \cdots, b_{mn}). \quad (10)$$

In the stationary state, their respective expected scores are the

$$\varpi_\alpha = \frac{v \cdot \omega_\alpha}{v \cdot 1} = \frac{D(p,q,\omega_\alpha)}{D(p,q,1)}, \quad (11)$$

$$\varpi_\beta = \frac{v \cdot \omega_\beta}{v \cdot 1} = \frac{D(p,q,\omega_\beta)}{D(p,q,1)}, \quad (12)$$

where **1** is the vector with all components 1. The denominators are needed because $v$ has not previously been normalized to have its components sum to 1 (as required for a stationary probability vector).

Because the scores in Eq. (11) and Eq.(12) depend linearly on their corresponding payoff matrices $P$, the same is true for any linear combination of scores, giving

$$a\varpi_\alpha + b\varpi_\beta + c = \frac{D(p,q,a\omega_\alpha + b\omega_\beta + c\mathbf{1})}{D(p,q,\mathbf{1})}. \qquad (13)$$

where $a$, $b$ and $c$ are constants.

This equation (13) reveals the possible linear relationship between the players' expected payoffs. Recalling that in the matrix $P'$ there exists a column $\bar{p}$ totally determined by $p^{(1)}$, or there exists a column $\bar{q}$ totally determined by $q^{(1)}$. If $\alpha$ chooses a strategy that satisfies

$$\bar{p} = a\omega_\alpha + b\omega_\beta + c\mathbf{1}, \qquad (14)$$

or if $\beta$ chooses a strategy with

$$\bar{q} = a\omega_\alpha + b\omega_\beta + c\mathbf{1}, \qquad (15)$$

then the determinant vanishes and a linear relation between the two expected scores,

$$a\varpi_\alpha + b\varpi_\beta + c = 0, \qquad (16)$$

will be imposed. Since matrix $P'$ is singular, the strategy $p$ which leads to the above linear Eq. (16) is a multi-strategy zero-determinant strategy of player $\alpha$.

**Feasible condition.**  Not all zero-determinant strategies are feasible, with probabilities $p$ and $q$ all in the range [0,1]. Whether they are feasible in any particular instance depends on the particulars of the application. A sufficient condition of feasible zero determinant strategies of multi-strategy games is as follows：$\sum_{i=1}^{nm} P_{inm} \neq 0$, and either $P_{inm} \geq 0, i = 1, 2, \cdots, nm$ or $P_{inm} \leq 0, i = 1, 2, \cdots, nm$.

**$\alpha$ Unilaterally Sets $\beta$'s Score.**  Eq. (13) allows much mischief by choosing values $a, b, c$ that keep her strategy $p$ as defined in Eq. (14) in the realm of possibility vectors, $\alpha$ can unilaterally enforce certain constraints on the iterated game's expected scores. From the above, $\alpha$ may choose to set $a = 0$, yielding $\varpi_\beta = -\frac{c}{b}$. By doing so, she can unilaterally determine $\beta$'s expected payoff.

$\alpha$ **Demands and Gets an Extortionate Share**. Interestingly, $\alpha$ may enforce a linear relation between her and $\beta$'s scores: $\alpha$ may ensure herself a multiple of every surplus $\beta$ earns over a certain offset. By setting

$$c = -(a+b)\Delta, \text{ for any offset } \Delta, \tag{17}$$

$\alpha$ enforces

$$\varpi_\alpha - \Delta = \lambda(\varpi_\beta - \Delta). \tag{18}$$

For values $\lambda > 1$ such strategies could be described as enforcing an "unfair", extortionate share of payoffs for $\alpha$.

## 3 Zero-determinant strategies of symmetric games

The definition of symmetric games is as follows: The payoff matrix of player $\alpha$ is $A = [a_{ij}]_{n \times n}$, while the payoff matrix of palyer $\beta$ is the transpose of $A$, that is, $B = A^T = [a_{ji}]_{n \times n}$, then

$$(A, B) = (a_{ij}, a_{ji})_{n \times n} = \begin{pmatrix} (a_{11}, a_{11}) & (a_{12}, a_{21}) & (a_{13}, a_{31}) & \cdots (a_{1n}, a_{n1}) \\ (a_{21}, a_{12}) & (a_{22}, a_{22}) & (a_{23}, a_{32}) & \cdots (a_{2n}, a_{n2}) \\ (a_{31}, a_{13}) & (a_{32}, a_{23}) & (a_{33}, a_{33}) & \vdots (a_{3n}, a_{n3}) \\ \cdots & \cdots & \cdots & \cdots & \cdots \\ (a_{n1}, a_{1n}) & (a_{n2}, a_{2n}) & (a_{32}, a_{23}) & \cdots (a_{nn}, a_{nn}) \end{pmatrix}. \tag{19}$$

If we assume that $(a_{ij}, a_{ji}) = (a_{ji}, a_{ij}), i, j = 1, 2, \cdots, n$, then matrix $(A, B) = (a_{ij}, a_{ji})_{n \times n}$ is symmetric. This is also our reason for calling it a symmetric game. For example, the prisoners' dilemma, stag hunt and the game of chicken are symmetric games, which means that they can all be represented in a symmetric $2 \times 2$ payoff matrix (see Eq.(20))

$$\begin{pmatrix} (R, R) & (S, T) \\ (T, S) & (P, P) \end{pmatrix}. \tag{20}$$

Let

$$(p_1, p_2, \cdots, p_n, p_{n+1}, p_{n+2}, \cdots, p_{2n}, \cdots, p_{(n-1)n+1}, p_{(n-1)n+2}, \cdots, p_{nn})$$
$$= (p^{(1)}_{\alpha_1\beta_1}, p^{(1)}_{\alpha_1\beta_2}, \cdots, p^{(1)}_{\alpha_1\beta_n}, p^{(1)}_{\alpha_2\beta_1}, p^{(1)}_{\alpha_2\beta_2}, \cdots, p^{(1)}_{\alpha_2\beta_n}, \cdots, p^{(1)}_{\alpha_n\beta_1}, p^{(1)}_{\alpha_n\beta_2}, \cdots, p^{(1)}_{\alpha_n\beta_n}),$$

$$(q_1, q_2, \cdots, q_n, q_{n+1}, q_{n+2}, \cdots, q_{2n}, \cdots, q_{(n-1)n+1}, q_{(n-1)n+2}, \cdots, q_{nn})$$
$$= (q^{(1)}_{\beta_1\alpha_1}, q^{(1)}_{\beta_1\alpha_2}, \cdots, q^{(1)}_{\beta_1\alpha_n}, q^{(1)}_{\beta_2\alpha_1}, q^{(1)}_{\beta_2\alpha_2}, \cdots, q^{(1)}_{\beta_2\alpha_n}, \cdots, q^{(1)}_{\beta_n\alpha_1}, q^{(1)}_{\beta_n\alpha_2}, \cdots, q^{(1)}_{\beta_n\alpha_n}).$$

We rewrite the payoff matrix player $\alpha$ as follows vector form

$$\omega_\alpha = (a_{11}, a_{12}, \cdots, a_{1n}, a_{21}, a_{22}, \cdots, a_{2n}, \cdots, a_{n1}, a_{n2}, \cdots, a_{nn}), \tag{21}$$

We rewrite the payoff matrix player $\beta$ as follows vector form

$$\omega_\beta = (a_{11}, a_{21}, \cdots a_{n1}, a_{12}, a_{22}, \cdots, a_{n2}, \cdots, a_{1n}, a_{2n} \cdots, a_{nn}). \tag{22}$$

Without loss of generality, assuming that $a_{11} \geq a_{nn}$, game $(\alpha_1, \beta_1)$ is mutual full cooperation, and game $(\alpha_n, \beta_n)$ is full noncooperation.

**Mischief and extortion**.  Extortion factor $\lambda$: if there exist constant $\lambda \geq 1$, such that,

$$a_{1j} - a_{nn} - \lambda(a_{1j} - a_{nn}) \leq 0, \ j = 2, 3, \cdots, n, \tag{23}$$

$$a_{ij} - a_{nn} - \lambda(a_{ji} - a_{nn}) \geq 0, \ i = 2, 3, \cdots, n-1, \ j = 1, 2, \cdots, n, \tag{24}$$

$$a_{nj} - a_{nn} - \lambda(a_{jn} - a_{nn}) \geq 0, \ j = 1, 2, \cdots, n-1, \tag{25}$$

Then factor $\lambda$ is called an extortion factor of this game.

Next, what if player $\alpha$ attempts to enforce an extortionate share of payoffs larger than the mutual full noncooperation value $a_{nn}$? Player $\alpha$ can do this by choosing

$$\bar{p} = \theta[(\omega_\alpha - a_{nn}\mathbf{1}) - \lambda(\omega_\beta - a_{nn}\mathbf{1})], \tag{26}$$

where $\lambda \geq 1$ is the extortion factor. Solving these equations for the *p*'s gives

$$\begin{aligned}
p_1 &= 1 - \theta(\lambda - 1)(a_{11} - a_{nn}) \\
p_2 &= 1 + \theta[a_{12} - a_{nn} - \lambda(a_{21} - a_{nn})] \\
&\vdots \\
p_n &= 1 + \theta[a_{1n} - a_{nn} - \lambda(a_{n1} - a_{nn})] \\
p_{n+1} &= \theta[a_{21} - a_{nn} - \lambda(a_{12} - a_{nn})] \\
p_{n+2} &= \theta[a_{22} - a_{nn} - \lambda(a_{22} - a_{nn})] \\
&\vdots \\
p_{2n} &= \theta[a_{2n} - a_{nn} - \lambda(a_{n2} - a_{nn})] \\
&\vdots \\
p_{(n-1)n+1} &= \theta[a_{n1} - a_{nn} - \lambda(a_{1n} - a_{nn})] \\
p_{(n-1)n+2} &= \theta[a_{n2} - a_{nn} - \lambda(a_{2n} - a_{nn})] \\
&\vdots \\
p_{(n-1)n+n-1} &= \theta[a_{nn-1} - a_{nn} - \lambda(a_{n-1n} - a_{nn})] \\
p_{nn} &= 0
\end{aligned} \tag{27}$$

Evidently, feasible strategies exist for extortion factor $\lambda$ and sufficiently small $\theta$.

If $n = 2$, from Eq.(23), Eq.(24) and Eq.(25), extortion factor $\lambda \geq 1$, such that,

$$a_{12} - a_{22} - \lambda(a_{21} - a_{22}) \leq 0, \tag{28}$$

$$a_{21} - a_{22} - \lambda(a_{12} - a_{22}) \geq 0. \tag{29}$$

From Eq.(27), we see that, for $n = 2$,

$$\begin{aligned} p_1 &= 1 - \theta(\lambda - 1)(a_{11} - a_{22}) \\ p_2 &= 1 + \theta[a_{12} - a_{22} - \lambda(a_{21} - a_{22})] \\ p_3 &= \theta[a_{21} - a_{22} - \lambda(a_{12} - a_{22})] \\ p_{22} &= 0 \end{aligned} \quad (30)$$

The Snowdrift game, the Hawk–Dove game and the Chicken game can reduce to the following symmetric game with payoff matrix

$$\begin{pmatrix} 1 & 1-r \\ 1+r & 0 \end{pmatrix}, \quad (31)$$

where $r$ is called the profit and loss ratio. From Eq.(28) and Eq.(29), if $r \geq 1$, for any real number $\lambda \geq 1$, it is an extortion factor; If $0 < r < 1$, the extortion factor $\lambda$ satisfies $1 \leq \lambda \leq \frac{1+r}{1-r}$. Now Eq.(30) can be shown that

$$\begin{aligned} p_1 &= 1 - \theta(\lambda - 1) \\ p_2 &= 1 - \theta[(\lambda + 1)r + \lambda - 1] \\ p_3 &= \theta[1 + r - \lambda(1 - r)] \\ p_{22} &= 0 \end{aligned} \quad (32)$$

From Eq.(18), we know that

$$\varpi_\alpha = \lambda \varpi_\beta. \quad (33)$$

Under the extortionate strategy, although $\alpha$'s score depends on $\beta$'s strategy $q$, the average payoff of $\alpha$ is $\lambda$ times the average payoff of $\beta$.

## 4 Conclusions

Press and Dyson have fundamentally changed the viewpoint on the Prisoner's Dilemma. Although their discovery makes us both excited and worried since a selfish person seems to have a more powerful mathematical tool to extort payoffs from those kindhearted and simpleminded people, it has influenced the way we think about the world. The study of zero-determinant strategies enables exciting new perspectives in the study of iterated multi-strategy games. The paper develops a theory for ZD strategies for multi-strategy games, with any number of strategies. We give a feasible condition of ZD strategies in multi-strategy games and a feasible extortion factor of symmetric games. The analytical results exhibit a similar yet different scenario to the case of two-strategy games. The results are also applied to the Snowdrift game, the Hawk–Dove game and the Chicken game. Zero-determinant strategies in iterated Prisoner's Dilemma can be seen as degenerate case of our results.

## ACKNOWLEDGMENTS

The authors acknowledge support from the Shanghai First-class Academic Discipline Project, China (Grant No. S1201YLXK), and supported by the Hujiang Foundation of China (Grant No. A14006).

**References**

[1] Stewart A J, Plotkin J B (2012) Extortion and cooperation in the Prisoner's Dilemma. *Proc Natl Acad Sci USA* 109: 10134–10135.

[2] Press W H, Dyson F J (2012) Iterated Prisoner's Dilemma contains strategies that dominate any evolutionary opponent. *Proc Natl Acad Sci USA* 109:10409–10413.

[3] Szolnoki A, Perc M (2014) Defection and extortion as unexpected catalysts of unconditional cooperation in structured populations. *Scientific Reports* 4: 5496

[4] Hilbe C, Rohl T, Milinski M (2014) Extortion subdues human players but is finally punished in the prisoner's dilemma. *Nature Communications* 5: 3976

[5] Stewart A J, Plotkin J B (2013) From extortion to generosity, evolution in the Iterated Prisoner's Dilemma. *Proc Natl Acad Sci USA* 110: 15348-15353.

[6] Adami C, Hintze A (2013) Evolutionary instability of zero-determinant strategies demonstrates that winning is not everything. *Nature Communications* 4: 2193

[7] Hao D, Rong Z H, Zhou T (2014) Zero-determinant strategy: An underway revolution in game theory. *Chin. Phys. B* 23(7): 078905

[8] Hilbe C, Traulsen A, Wu B, Nowak M A (2014) Zero-determinant alliances in multiplayer social dilemmas. arXiv:1404.2886v1

[9] Pan L, Hao D, Rong Z, and Zhou T (2014) Zero-Determinant Strategies in the Iterated Public Goods Game. arXiv:1402.3542v1.

[10] Roemheld L (2013) Evolutionary Extortion and Mischief: Zero Determinant strategies in iterated 2x2 games. arXiv:1308.2576v1.